# Towards ML Engineering: A Brief History Of TensorFlow Extended (TFX)

Konstantinos (Gus) Katsiapis, Abhijit Karmarkar, Ahmet Altay, Aleksandr Zaks, Neoklis Polyzotis, Anusha Ramesh, Ben Mathes, Gautam Vasudevan, Irene Giannoumis, Jarek Wilkiewicz, Jiri Simsa, Justin Hong, Mitch Trott, Noé Lutz, Pavel A. Dournov, Robert Crowe, Sarah Sirajuddin, Tris Brian Warkentin, Zhitao Li

Google Inc.

```
{katsiapis, awk, altay, azaks, npolyzotis, anusharamesh, benmathes,
   gvasudevan, irenegi, jarekw, jsimsa, hongjustin, trott, noelutz,
dournov, robertcrowe, ssirajuddin, triswarkentin, zhitaoli}@google.com
```

## Abstract

Software Engineering, as a discipline, has matured over the past 5+ decades. The modern world heavily depends on it, so the increased maturity of Software Engineering was an eventuality. Practices like testing and reliable technologies help make Software Engineering reliable enough to build industries upon. Meanwhile, Machine Learning (ML) has also grown over the past 2+ decades. ML is used more and more for research, experimentation and production workloads. ML now commonly powers widely-used products integral to our lives.

But ML Engineering, *as a discipline*, has not widely matured as much as its Software Engineering ancestor. Can we take what we have learned and help the nascent field of applied ML evolve into ML Engineering the way Programming evolved into Software Engineering [1]?

In this article we will give a whirlwind tour of Sibyl [2] and TensorFlow Extended (TFX) [3], two successive end-to-end (E2E) ML platforms at Alphabet. We will share the lessons learned from over a decade of applied ML built on these platforms, explain both their similarities and their differences, and expand on the shifts (both mental and technical) that helped us on our journey. In addition, we will highlight some of the capabilities of TFX that help realize several aspects of ML Engineering. We argue that in order to unlock the gains ML can bring, organizations should advance the maturity of their ML teams by investing in robust ML infrastructure and promoting ML Engineering education. We also recommend that before focusing on cutting-edge ML modeling techniques, product leaders should invest more time in adopting interoperable ML platforms for their organizations. In closing, we will also share a glimpse into the future of TFX.

## Where We Are Coming From

Applied ML has been an integral part of Google products and services over the last decade, and is becoming more so over time. We discovered early on from our endeavors to apply ML in production that while ML algorithms are important, they are usually insufficient in realizing the successful application of ML in a product [4]. In particular, E2E ML platforms, which help with all aspects of the ML lifecycle, are usually need-ed to both accelerate ML adoption and make its use durable and sustainable.

## Sibyl (2007 - 2020)

E2E ML platforms are not a new thing at Google. Sibyl [2], founded in 2007, was a platform that enabled massive-scale ML, catered to production use. Sibyl offered a decent amount of modeling flexibility on top of "wide" models (linear, logistic, poisson regression and later factorization machines [5]) coupled with non-linear transformations and customizable loss functions and regularization [6]. Importantly, Sibyl also offered tools for several aspects of the ML workflow including Data Ingestion, Data Analysis and Validation, Training (of course), Model Analysis, and Training-Serving Skew Detection. All these were packaged as a single integrated product that allowed for iterative experimentation. This holistic product offering, coupled with the Sibyl team's user focus, rendered Sibyl to, once upon a time, be one of the most widely used E2E ML platforms at Google. Sibyl has since been decommissioned. It was in production for ~14 years, and the vast majority of its workloads migrated to TFX.

## TFX (2017 - ?)

While several of us were still working on Sibyl, a notable revolution was happening in the ML algorithms fields with the popularization of Deep Learning (DL). In 2015, Google publicly released TensorFlow [7] (which was itself a successor to a previous system called DistBelief [8]). Since its inception, TensorFlow supported a variety of applications with a focus on DL training and inference. Its flexible programming model allowed it to be used for a lot more than DL and its popularity in both research and production positioned it as the lingua franca for authoring ML algorithms. While TensorFlow offered flexibility, it lacked a complete end-to-end production system. On the other hand, Sibyl had robust end-to-end capabilities, but lacked flexibility. It became apparent that we needed an E2E ML platform for TensorFlow in order to accelerate ML at Google; in 2017, nearly a decade after the birth of Sibyl, we launched TFX within Google. TFX is now the most widely used, general purpose E2E ML platform at Alphabet, including Google.

In the 3 years since its launch, TFX has enabled Alphabet to realize what might be described as "industrial-scale" ML: TFX is used by thousands of users within Alphabet, and it powers hundreds of popular Alphabet products, including Cloud AI services on Google Cloud Platform (GCP). On any given day there are thousands of TFX pipelines running, which are processing exabytes of data and producing tens of thousands of models, which in turn are performing hundreds of millions of inferences per second. TFX's widespread adoption helps Alphabet realize the flow of research into production and enables very diverse use cases for both direct and indirect TFX users. This widespread adoption also enables teams to focus on model development rather than ML platform development, allowing ML to be more easily used in novel product areas, and creating a virtuous cycle of ML platform evolution from ML applications.

The popularity and impact of TensorFlow [9] within and outside of Alphabet, the popularity and impact of TFX within Alphabet, and the reality that equivalents of ML engineering will be needed by organizations and individuals everywhere in the world, felt like something we could not ignore. That led us to publicly describe the design and initial deployment of TFX within Google [10] and to, step by step, make more of our learnings and our technology publicly available (including open source), while we continue building more of each. We were able to accomplish this in part because, like Sibyl, TFX built upon robust infrastructural dependencies. For example, Sibyl made heavy use of MapReduce [11] and its successor Flume [12] for its distributed data processing, and now TFX heavily uses their portable successor, Apache Beam [13], for the same.

Following in TensorFlow's footsteps, the public TFX offering [3] was released in early 2019 and widely adopted in under a year across environments including on-premises and GCP with Cloud AI Platform Pipelines [14]. Some of our partners have also publicly shared their use cases powered by TFX [15], including how it radically improved their applied ML velocity [16].

# Lessons From Our 10+ Year Journey Of ML Platform Evolution

Though the journey of ML Platform(s) evolution at Google has been a long and exciting one, we expect that the majority of excitement is yet to come! To that end, we want to share a summary of our learnings, some of which were more painfully gained than others. The learnings fall into two categories, namely what remained the same as part of the evolution, but also what changed, and why! We present the learnings in the context of two successive platforms, Sibyl and TFX, though we believe them to be widely applicable.

## What Remains The Same And Why

The areas discussed in this section capture a few examples of things that seem enduring and pass the test of time. As such, we expect these to also remain applicable in the future, across different incarnations of ML platforms and frameworks. We look at these from both an applied ML perspective and an infrastructure perspective.

### Applied ML

#### The Rules Of Machine Learning

Successfully applying ML to a product is very much a discipline. It involves a steep learning curve and necessitates some mental model shifts (or perhaps augmentations). To make this challenging task easier, we have publicly shared *The Rules of Machine Learning* [17, 18]. These are rules that represent learnings from iteratively applying ML to a lot of products at Google. Notably, the adoption of ML in Google products illustrates a common evolution:

- Start with simple rules and heuristics, and generate data to learn from; this journey usually starts from the serving side [17].
- Move to simple ML (i.e., simple models) and realize large gains; this is usually the entry point for introduction of ML pipelines [17].
- Move to ML with more features and more advanced models to realize decent gains [17].
- Move to state-of-the-art ML, manage refinement and complexity (for solutions to the problems that are worth it), and realize small gains [17].
- Apply the above launch-and-iterate cycle to more aspects of products and to solve more problems, bearing in mind return on investment (and diminishing returns).

We have found *The Rules of Machine Learning* to be steadfast across platforms and time and we hope they end up being as valuable to others as they have been to us and our users. In particular, we believe that following the rules will help others be better at the discipline of ML engineering, including helping them avoid the mistakes that we and our users have made in the past. TFX is an attempt to codify these rules, quite literally, in code. We hope to benefit ourselves but also accelerate ML, done well, for the entire industry.

## The Discipline Of ML Engineering

In developing *The Rules of Machine Learning*, we realized that the discipline for building robust systems where the core logic is produced by complex processes involving both code and data requires additional scrutiny beyond that which software engineering provides. As such, we define *ML Engineering* as a superset of the discipline of software engineering designed to handle the unique complexities of the practical application of ML.

Attempting to summarize the totality of the discipline of ML engineering would be somewhat difficult, if not impossible, especially given how our understanding of it is still limited, and the discipline itself continues to evolve. We do take solace in the following though:

- The limited understanding we do have seems to be enduring across platforms and time.
- Analogy can be a powerful tool, so several aspects of the better understood discipline of software engineering have helped us draw parallels of how ML engineering could evolve from ML programming, much like how software engineering evolved from programming [1].

An early realization we had was the following: artifacts are first class citizens in ML, on par with the processes that produce and consume them.

This realization affected the implementation and evolution of Sibyl; it was entrenched in TFX by the time we publicly wrote about it [10] and was ultimately generalized and formalized in ML Metadata [19], now powering TFX.

Below we present fundamental elements of ML engineering, some examples of ML artifacts and their first class citizenship, and make an attempt to draw analogies with software engineering where possible.

### Data

Similarly to how code is at the heart of software, data is at the heart of ML. Data management represents serious challenges in production ML [20]. Perhaps the simplest analogy would be to think about what constitutes a unit test for data. Unit tests verify expectations on how code should behave, by testing the contracts of the pertinent code and instilling trustworthiness in said contracts. Similarly, setting explicit expectations on the form of the data (including its schema, invariants and value distributions), and checking that the data agrees with implicit expectations embedded in the training code can, more so together, make the data trustworthy enough to train models on [21]. Though unit tests can be exhaustive and verify strong contracts, data contracts are in general a lot weaker even if they are necessary. Though unit tests can be exhaustively consumed and verified by humans, data can usually be meaningful to humans only in summarized fashion.

Just as code repositories and version control are pillars for managing code evolution in software engineering, systems for managing data evolution and understanding are pillars of ML engineering.

TFX's ExampleGen [22], StatisticsGen [22], SchemaGen [22] and ExampleValidator [22] components help one treat data as first class citizens, by enabling data management, analysis and validation in (continuous) ML pipelines [23].

### Models

Similarly to how a software engineer produces code that is compiled into programs, an ML engineer produces data and code which is "compiled" into ML programs, more commonly known as models. These two kinds of programs are however very different in nature. Though programs that come out of software usually have strong contracts, models have much weaker contracts. These weak contracts are usually statistical in nature and as such only verifiable in some summarized form (such as a model having sufficient accuracy on a subset of labeled data). This is not at all surprising since models are the product of code and data, and the latter itself doesn't have strong contracts and is also only digestible in summarized form.

Just as code and data evolve over time, models also evolve over time. However, model evolution is more complicated than the evolution of its constituent code and data. For example, high test coverage (with fuzzing) can give good confidence in both the correctness and the correct evolution of a piece of code, but out-of-distribution and counterfactual yet realistic data for model evaluation can be notoriously difficult to produce.

In the same way that putting together multiple programs in a system necessitates integration testing which is a pillar of software engineering, putting together code and data necessitates end-to-end model validation and understanding [24] which is a pillar of ML engineering.

TFX's Evaluator [22] and InfraValidator [22] components provide validation and understanding of models, treating them as first class citizens of ML engineering.

### Mergeable Fragments

Similarly to how a software engineer merges together pre-existing libraries (or systems) with their code in order to build useful programs, an ML engineer merges together code fragments, data fragments, analysis fragments and model fragments on a regular basis in order to build useful ML pipelines. A notable difference between software engineering and ML engineering is that even when the code is fixed for the latter, data is usually volatile for it (e.g. new data arrives on a regular basis) and as such the downstream artifacts need to be produced frequently and efficiently. For example, a new version of a model usually needs to be produced if any part of its input data has changed. As such, it is important for ML pipelines to produce artifacts that are mergeable. For example, a summary of statistics from one dataset should be easily mergeable with that of another dataset such that it is easy to summarize the statistics of the union of the two datasets. Similarly, it should be easy to transfer the learnings of one model to another model in general, and the learnings of a previous version of a model to the next version of the same model in particular.

There is however a catch, which relates to the previous discussion regarding the equivalents of test coverage for models. Merging new fragments into a model could necessitate creation of novel out-of-distribution and counterfactual evaluation data, contributing to the difficulty of (efficient) model evolution, thus rendering it a lot harder than pure code evolution.

TFX's ExampleGen [22], Transform [22], Trainer [22] and Tuner [22] components, together with TensorFlow Hub [25], help one treat artifacts as first class citizens by enabling production and consumption of mergeable fragments in workflows that perform data caching, analyzer caching, warmstarting and transfer learning.

### Artifact Lineage

Despite all the advanced methodology and tooling that exists for software engineering, the programs and systems that are built invariably need to be debugged. The same holds for ML programs, but debugging them is notoriously harder because non-proximal effects are a lot more prevalent for ML programs due to the plethora of artifacts involved. A model might be inaccurate due to bad artifacts from several sources of error, including flaws in the code, the learning algorithm, the training data, the serving path, or the serving data, to name a few. Much like how stack traces are invaluable for identifying root causes of defects in software programs, the lineage of all artifacts produced and consumed by an ML pipeline is invaluable for identifying root causes of defects in ML models. Additionally, by knowing which downstream artifacts were produced from a problematic artifact, we can identify all impacted systems and users and take mitigating actions.

TFX's use of ML Metadata (MLMD) [19] helps treat artifacts as first class citizens. MLMD enables advanced cataloging and querying of metadata and lineage associated with artifacts which can together increase the confidence of sharing artifacts even outside the boundaries of a pipeline. MLMD also helps with advanced debugging and, when coupled with the underlying data storage layer, forms the foundation of TFX's ML compliance mechanisms.

### Continuous Learning And Unlearning

ML production pipelines operate in a dynamic environment:

- New data can arrive continuously.
- The modeling code can change, particularly in the early stages of model development.
- The surrounding infrastructure can change, e.g., a new version of some underlying (ML) library.

When changes happen, a pipeline needs to react, often by rerunning its steps in the new environment. This dynamicity increases the importance of provenance tracking in order to facilitate debugging and root-cause analysis. As a simple example, to debug a model failure, it is necessary to know not only which data was used to train the model, but also the versions of the modeling code and any surrounding infrastructure.

ML pipelines must also support low-friction mechanisms to handle these changes. Consider for example the arrival of new data, which necessitates retraining the model. This is a natural requirement in rapidly changing environments, like recommender systems or adversarial systems. Requiring the user to manually retrain the model can be unrealistic, given that the data can arrive at a regular and frequent rate. Instead, we can employ automation by way of "continuous training", where the pipeline detects the presence of new data and automatically schedules the generation of updated models. In turn, this functionality requires automatically: orchestrating work based on the presence of artifacts (including data), recovering from intermittent failures, and catching up to real-time when recovering [26]. It is common for ML pipelines to run for years ingesting code and data, continuously producing models that make predictions that inform decisions.

Another example of a low-friction mechanism is support for "backfilling" an ML pipeline. In this case, the user might need to rerun the pipeline on existing artifacts but using updated versions of the components, such as rerunning the trainer on existing data using a new version of the modeling code/library. Another use of backfilling is rerunning the pipeline with new versions of existing data, say, to fix an error in the data. These backfills are orthogonal to continuous training and can be used together. For instance, the user can manually trigger a rerun of the trainer, and the generated model artifact can then automatically trigger model evaluation and validation.

TFX was built from the ground up in a way that enables continuous learning (and unlearning) which fundamentally shaped its design. At the same time, these advanced capabilities also allow it to be used in a "one-shot", discontinuous, fashion. In fact, within Alphabet, both modes of deployment are widely used. Moreover, TFX also supports different types of backfill operations to enable fine-grained interventions during normal pipeline execution.

Even though the public TFX offering [22] doesn't yet offer continuous ML pipelines, we are actively working on making our existing technology portable so that it can be made publicly available [e.g 27].

## Infrastructure

### Building On The Shoulders Of Giants

Realizing ambitious goals necessitates building on top of solid foundations, collaborating with others and leveraging each other's work. TFX reuses many of Sibyl's system designs, hardened over a decade of Sibyl's production ML experience. Additionally, TFX incorporates new technologies in areas where robust standards emerged:

- Similarly to how Sibyl built its algorithms and workflows on top of MapReduce, TFX leverages both TensorFlow [9] and Apache Beam [13] for its distributed training and data processing workflows.
- Similarly to how Sibyl was columnar, TFX adopted Apache Arrow [28] as the columnar in-memory representation for its compute-intensive libraries.

Taking dependencies where robust standards have emerged has allowed TFX and its users to achieve seamless performance and scalability. It also enables TFX to focus its energy on building the deltas of what is needed for applied ML, as opposed to re-implementing difficult-to-get-right technology. Some of our dependencies, like Kubeflow Pipelines [29] or Apache Airflow [30], are selected by TFX's users themselves when the value / features they get from them outweigh the costs that the additional dependencies entail.

Taking dependencies unfortunately incurs costs. We have found that taking dependencies requires effort that is super-linear to the number of dependencies. Said costs are often absorbed by us and our sister teams but can (and sometimes do) leak to our users, usually in the form of conflicting (version) dependencies or incompatibilities between environments and dependencies.

### Interoperability And Positive Externalities

ML platforms do not operate in a vacuum. They instead operate within the context of a bigger system or infrastructure, connecting to data producing sources upstream and model consuming sinks downstream, which in turn frequently produce the data that feeds the ML platform, thereby closing the loop. Strong adoption of a platform usually necessitates interoperability with other important technologies in its environment.

- Similarly to how Sibyl interoperated with Google's Ads technology stack for data ingestion and model serving, TFX offers a plethora of connectors for data ingestion and allows serving the produced model in multiple deployment environments and devices.
- Similarly to how Sibyl interoperated with Google's compute stack, TFX leverages Apache Beam [13] to execute on Apache Flink [31] and Apache Spark [32] clusters as well as serverless offerings like Google Cloud Dataflow [33].
- TFX built an orchestration abstraction on top of MLMD [19] and provides orchestration options on top of Apache Airflow [22], Apache Beam [22], Kubeflow Pipelines [22] as well as the primitives to integrate with one's custom orchestrator. MLMD itself works with several relational databases like SQLite [34] and MySQL [35].

Interoperability necessitates some amount of abstraction and standardization and usually enables sum-greater-than-its-parts effects. TFX is both a beneficiary and a benefactor of the positive externalities created by said interoperability, both within and outside of Alphabet. TFX's users are also beneficiaries of the interoperability as they can more easily deploy and use TFX on top of their existing installed base.

Interoperability also comes with costs. The combination of multiple technology stacks can lead to an exponential number of distinct deployment configurations. While we test some of the distinct deployment configurations end-to-end and at-scale, like for example TFX on GCP [14], we have neither the expertise nor the resources to do so for the combinatorial explosion of all possible deployment options. We thus encourage the community to work with us on the deployment configurations that are most useful for them [36].

# What Is Different And Why

The areas discussed in this section capture a few examples of things that needed to change in order for our ML platform to adapt to a new reality and as such remain useful and impactful.

## Environment And Device Portability

Sibyl was a massive scale ML platform designed to be deployed on Google's large-scale cluster, namely Borg [37]. This made sense as applied ML at Google was, originally, primarily used in products that were widely used. As ML expertise grew across the world, and ML could be applied to more use cases (large and small) across environments both within and outside of Google, the need for portability gradually but surely became a hard constraint.

- While Sibyl ran only on Google's datacenters, TFX runs on laptops, workstations, servers, datacenters, and public Clouds. In particular, when TFX runs on Google's Cloud [38], it leverages automation and optimizations offered by GCP Services, enabled by Google's unique infrastructure.
- While Sibyl ran only on CPUs, TFX leverages TensorFlow to run on different kinds of hardware including CPUs, GPUs and Google's TPUs [39, 40].
- While Sibyl's models ran on servers, TFX leverages TensorFlow to produce models that run on laptops, workstations, and servers via TensorFlow Serving [22, 41] and Apache Beam [22], on mobile and IoT devices via TensorFlow Lite [42], and on browsers via TensorFlow JS [43].

TFX's portability enabled it to be used in a very diverse set of environments and devices, in order to solve problems from small scale to massive scale.

Unfortunately, portability comes with costs. We have found that maintaining a portable core with environment-specific and device-specific specialization requires effort that is super-linear to the number of environments / devices. Said costs are however largely absorbed by us and our sister teams and as such are frequently not visible to our users.

## Modularity And Layering

Even though Sibyl's offering as an integrated product was immensely valuable, its structure and interface were somewhat monolithic, limiting it to a specific set of "direct" users who would have to adopt it wholesale. In contrast, TFX evolved to be a modular and layered architecture, and became more so over time as partnerships with other teams and products grew. Notable layers (with examples) in TFX include:

| Layer | Examples |
|---|---|
| ML Services | - Cloud AutoML [44]<br>- Cloud Recommendations AI [45]<br>- Cloud AI Platform [14, 22, 46]<br>- Cloud Dataflow [33, 47]<br>- Cloud BigQuery [48] |
| Pipelines (of composable Components) | - TensorFlow Extended (TFX) [3, 22, 49] |

| Binaries | • TensorFlow Serving (TFS) [41, 50] |
|---|---|
| Libraries | • TensorFlow Data Validation (TFDV) [51]<br>• TensorFlow Transform (TFT) [52]<br>• TensorFlow Hub (TFH) [53]<br>• TensorFlow Model Analysis (TFMA) [54]<br>• TFX Basic Shared Libraries (TFX_B-SL) [55]<br>• ML Metadata (MLMD) [56] |

TFX's layered architecture enables it to be used by a very diverse set of users whether that's piecemeal via its libraries, wholesale via its pipelines (with or without the pertinent services), or in a fashion that's completely oblivious to the end users (e.g. by them using ML services which TFX powers under the hood)!

Unfortunately, layering comes with costs. We have found that maintaining multiple publicly accessible layers of our product requires effort that is roughly linear to the number of layers. Said costs occasionally leak to our users in the form of confusion regarding what layer makes the most sense for them to use.

## Multi-faceted Flexibility

Even though Sibyl was more flexible in terms of modeling capabilities compared to available alternatives at the time, aspects of its flexibility across several parts of the ML workflow fell short of Google's needs for accelerating ML for novel use cases, which led to the development of TFX [22].

- While Sibyl only offered specific kinds of data analysis, TFX's StatisticGen component [22] offers more built-in capabilities and the ability to realize custom analyses, via TensorFlow Data Validation [22].
- While Sibyl only offered transformations that were pure composable mappers, TFX's Transform component [22] offers more mappers, custom mappers, more analyzers, custom analyzers, as well as arbitrarily composed (custom) mappers and (custom) analyzers, via TensorFlow Transform [22].
- While Sibyl only offered "wide" models, TFX's Trainer component [22] offers any model that can be realized on top of TensorFlow [57], including models that can be shared and can transfer-learn, via TensorFlow Hub [25].
- While Sibyl only offered automatic feature crossing (a.k.a. feature conjunctions) on top of "wide" models, TFX's Tuner component [22] allows for arbitrary hyper parameter optimization based on state of the art algorithms.
- While Sibyl only offered specific kinds of model analysis, TFX's Evaluator component [22] offers more built-in metrics, custom metrics, confidence intervals and fairness indicators [58], via TensorFlow Model Analysis [22].
- While Sibyl's pipeline topology was fixed (albeit somewhat customizable), TFX's SDK allows one to create custom (optionally containerized) components [22] and use them together with standard components [22] in a flexible and fully customizable pipeline topology [22].

The increase of flexibility in all these dimensions enabled improved experimentation, wider reach, more use cases, as well as accelerated flow from research to production.

Flexibility does not come without costs. A more flexible system is one that is harder to get right in the first place as well as harder for us to main-tain and to evolve as developers of the ML platform. Users may also have to manage increased complexity as they take advantage of this flexibility. Furthermore, we might not be able to offer as strong of a support story on top of an ML platform that is Turing complete.

# Where We Are Going

Armed with the knowledge of the past, we present a glimpse of what we plan for the future of TFX as of 2020, based on our roadmap [59], requests for comments (RFCs) [60], and contribution guidelines [36]. We will continue our work on enabling ML Engineering in order to democratize applied ML, and help everyone practice responsible AI [61] and apply it in a fashion that upholds Google's AI Principles [62].

## Drive Interoperability And Standards

In order to meet the demand for the burgeoning variety of ML solutions, we will continue to increase our technology's interoperability. Our work on interoperability and standards as well as open-sourcing more of our technology, reflects our principle to *"be socially beneficial"* as well as to *"be made available for uses that accord with these principles"* by making it easier for everyone to follow these practices. As part of this mission, we will empower the industry to build advanced ML systems by open-sourcing more of our technology, and by standardizing ML artifacts and metadata. Some select examples of this work include:

- TFX Standardized Inputs [63].
- Advanced TFX DSL semantics [64], Data Model and IR [65].
- Standardization of ML artifacts and metadata.
- Standardization of distributed workloads on heterogeneous runtime environments.
- Inference on distributed and streaming models.
- Improvements to interoperability with mobile and edge ML deployments.
- Improvements for ML framework interoperability and artifact sharing.

## Increase Automation

Automation is the backbone of reliable production systems, and TFX is heavily invested in improving and expanding its use of automation. Our work in increased automation reflects our principles of helping make ML deployments *"be built and tested for safety"* and *"avoid creating or reinforcing unfair bias"*. Some upcoming efforts include a TFX Pipeline testing framework, automated model improvement in the TFX Tuner [66], auto-detecting surprising model behavior on multidimensional slices, facilitating automatic production of Model Cards [67, 68] and improving our training-serving skew detection capabilities. TFX on GCP will also continue driving requirements for new (and will better make use of existing) advanced automation features of pertinent services.

## Improve ML Understanding

ML understanding is an important aspect of deploying production ML, and TFX is well positioned to provide significant gains in this field. Our work on improving ML understanding reflects our principles to help *"avoid creating or reinforcing unfair bias"* and help make ML deployments *"be accountable to people"*. Critical to understanding is to be able to track the lineage of artifacts used to produce a model, an area TFX will continue to invest in. Improvements to TFX technologies like struct2tensor [69] will further enable training, serving, and analyzing models on structured data, thus allowing reasoning about models closer to the original data semantics. We also plan to utilize TFX as a vehicle to expand support for fairness evaluation, remediation, and documentation.

## Uphold High Standards And Best Practices

As a vehicle for amplification of ML technology, TFX must continue to *"uphold high standards of scientific excellence"* and promote best practices. The team will continue publishing scientific papers and conducting public outreach via our existing channels, as well as offer educational courses in partnership with established institutions. We will also improve trust in our model analysis tools using integrated uncertainty measures by, for example, enabling scalable computation of confidence intervals for model metrics, and we will improve our training-serving skew detection capabilities. It's also critical for research and production to be able to have reproducible ML artifacts, enabled by our work in precise provenance tracking for auditing and reproducing models. Also key is reproducibility of measurements, driven by efforts like NitroML, which will provide tooling for benchmarking AutoML pipelines [[70](#)].

Given that several of the areas where we expand our technology are new to us, we will make an effort to distinguish the battle-tested from the experimental aspects of our technology, in order to enable our users to confidently choose the set of capabilities that meet their desires and needs.

## Improve Tooling

Despite TFX providing tools for aspects of ML engineering and several phases of the ML lifecycle, we believe this is still a nascent area. While improving tooling is a natural fit for TFX, it also reflects our principle of helping ML deployments *"be made available for uses that accord with these principles"*, *"supporting scientific excellence,"* and being *"built and tested for safety"* .

One area of improvement is applying ML to the data itself, be it through sensing anomalies or finding patterns in data or enriching data with predictions from ML models. Making it easy to enrich large volumes of data (especially critical streaming data used for low-latency, high volume actions) has always been a challenge. Bringing TFX capabilities into data processing frameworks is our first step here. We have already made it possible to enrich streaming events with labels or make predictions in Apache Beam and, by extension, Cloud Dataflow. We plan to follow this work by leveraging pre-built models (served out of Cloud AI Pipelines and TensorFlow Serving) to make adding a new field in a distributed dataset representing predictions from streams of models trivially easy.

Furthermore, while there are many tools for detecting, discovering, and auditing ML workflows, there is still a need for automated (or assisted) mitigation of discovered issues, and we will invest in this area. For example, proactively predicting which pipeline runs won't result in better models based on the currently-executing pipeline, perhaps even before training, can significantly reduce time and resources spent on creating poor models.

# A Joint Journey

Building TFX and exploring the fundamentals of ML engineering was the cumulative effort of many people over many years. As we continue to make strides and further develop this field, it's important we recognize the collaborative effort of those who got us here.

Of course, it will take many more collaborations to drive the future of this field, and as such, we invite *you* to join us on this journey *"Towards ML Engineering"!*

## The TFX Team

The TFX project is realized via collaboration of multiple organizations within Google. Different organizations usually focus on different technology and product layers, though there is a lot of overlap on the portable parts of our technology. Overall we consider ourselves a single team and below we present an alphabetically sorted list of current TFX team members who are contributors to the ideation, research, design, implementation, execution, deployment, management, and advocacy (to name a few) aspects of TFX; they continue to inspire, help, teach, and challenge each other to advance our field:

*Abhijit Karmarkar, Adam Wood, Aleksandr Zaks, Alina Shinkarsky, Neoklis Polyzotis, Amy Jang, Amy McDonald Sandjideh, Amy Skerry-Ryan, Andrew Audibert, Andrew Brown, Andy Lou, Anh Tuan Nguyen, Anirudh Sriram, Anna Ukhanova, Anusha Ramesh, Archana Jain, Arun Venkatesan, Ashley Oldacre, Baishun Wu, Ben Mathes, Billy Lamberta, Chandni Shah, Chansoo Lee, Chao Xie, Charles Chen, Chi Chen, Chloe Chao, Christer Leusner, Christina Greer, Christina Sorokin, Chuan Yu Foo, CK Luk, Connie Huang, Daisy Wong, David Smalling, David Zats, Dayeong Lee, Dhruvesh Talati, Doojin Park, Elias Moradi, Emily Caveness, Eric Johnson, Evan Rosen, Florian Feldhaus, Gal Oshri, Gautam Vasudevan, Gene Huang, Goutham Bhat, Guanxin Qiao, Gus Katsiapis, Gus Martins, Haiming Bao, Huanming Fang, Hui Miao, Hyeonji Lee, Ian Nappier, Ihor Indyk, Irene Giannoumis, Jae Chung, Jan Pfeifer, Jarek Wilkiewicz, Jason Mayes, Jay Shi, Jiayi Zhao, Jingyu Shao, Jiri Simsa, Jiyong Jung, Joana Carrasqueira, Jocelyn Becker, Joe Liedtke, Jongbin Park, Jordan Grimstad, Josh Gordon, Josh Yellin, Jungshik Jang, Juram Park, Justin Hong, Karmel Allison, Kemal El Moujahid, Kenneth Yang, Khanh LeViet, Kostik Shtoyk, Lance Strait, Laurence Moroney, Li Lao, Liam Crawford, Magnus Hyttsten, Makoto Uchida, Manasi Joshi, Mani Varadarajan, Marcus Chang, Mark Daoust, Martin Wicke, Megha Malpani, Mehadi Hassen, Melissa Tang, Mia Roh, Mig Gerard, Mike Dreves, Mike Liang, Mingming Liu, Mingsheng Hong, Mitch Trott, Muyang Yu, Naveen Kumar, Ning Niu, Noah Hadfield-Menell, Noé Lutz, Nomi Felidae, Olga Wichrowska, Paige Bailey, Paul Suganthan, Pavel Dournov, Pedram Pejman, Peter Brandt, Priya Gupta, Quentin de Laroussilhe, Rachel Lim, Rajagopal Ananthanarayanan, Rene van de Veerdonk, Robert Crowe, Romina Datta, Ron Yang, Rose Liu, Ruoyu Liu, Sagi Perel, Sai Ganesh Bandiatmakuri, Sandeep Gupta, Sanjana Woonna, Sanjay Kumar Chotakur, Sarah Sirajuddin, Sheryl Luo, Shivam Jindal, Shohini Ghosh, Sina Chavoshi, Sydney Lin, Tanya Grunina, Thea Lamkin, Tianhao Qiu, Tim Davis, Tris Warkentin, Varshaa Naganathan, Vilobh Meshram, Volodya Shtenovych, Wei Wei, Wolff Dobson, Woohyun Han, Xiaodan Song, Yash Katariya, Yifan Mai, Yiming Zhang, Yuewei Na, Zhitao Li, Zhuo Peng, Zhuoshu Li, Ziqi Huang, Zoey Sun, Zohar Yahav*

Thank you, all!

## The TFX Team … Extended

Beyond the current TFX team members, there have been many collaborators both within and outside of Alphabet whose discussions, technology, as well as direct and indirect contributions, have materially influenced our journey. Below we present an alphabetically sorted list of these collaborators:

*Abdulrahman Salem, Ahmet Altay, Ajay Gopinathan, Alexandre Passos, Alexey Volkov, Anand Iyer, Andrew Bernard, Andrew Pritchard, Chary Aasuri, Chenkai Kuang, Chenyu Zhao, Chiu Yuen Koo, Chris Harris, Chris Olston, Christine Robson, Clemens Mewald, Corinna Cortes, Craig Chambers, Cyril Bortolato, D. Sculley, Daniel Duckworth, Daniel Golovin, David Soergel, Denis Baylor, Derek Murray, Devi Krishna, Ed Chi, Fangwei Li, Farhana Bandukwala, Gal Elidan, Gary Holt, George Roumpos, Glen Anderson, Greg Steuck, Grzegorz Czajkowski, Haakan Younes, Heng-Tze Cheng, Hossein Attar, Hubert Pham, Hussein Mehanna, Irene Cai, James L. Pine, James Pine, James Wu, Jeffrey Hetherly, Jelena Pjesivac-Grbovic, Jeremiah Harmsen, Jessie Zhu, Jiaxiao Zheng, Joe Lee, Jordan Soyke, Josh Cai, Judah Jacobson, Kaan Ege Ozgun, Kenny Song, Kester Tong, Kevin Haas, Kevin Serafini, Kiril Gorovoy, Kostik Steuck, Kristen LeFevre, Kyle Weaver, Kym Hines, Lana Webb, Lichan Hong, Lukasz Lew, Mark Omernick, Martin Zinkevich, Matthieu Monsch, Michel Adar, Michelle Tsai, Mike Gunter, Ming Zhong, Mohamed Hammad, Mona Attariyan, Mustafa Ispir, Neda Mirian, Nicholas Edelman, Noah Fiedel, Panagiotis Voulgaris, Paul Yang, Peter Dolan, Pushkar Joshi, Rajat Monga, Raz Mathias, Reiner Pope, Rezsa Farahani, Robert Bradshaw, Roberto Bayardo, Rohan Khot, Salem Haykal, Sam McVeety, Sammy Leong, Samuel*

*Ieong, Shahar Jamshy, Slaven Bilac, Sol Ma, Stan Jedrus, Steffen Rendle, Steven Hemingray, Steven Ross, Steven Whang, Sudip Roy, Sukriti Ramesh, Susan Shannon, Tal Shaked, Tushar Chandra, Tyler Akidau, Venkat Basker, Vic Liu, Vinu Rajashekhar, Xin Zhang, Yan Zhu, Yaxin Liu, Younghee Kwon, Yury Bychenkov, Zhenyu Tan*

Thank you, all!